\documentclass[11pt,a4paper]{article}
\usepackage{amsmath}
\usepackage{amssymb}

\title{A proposal for the role of higher spin fields in Conformal Cyclic Cosmology}
\author{Anders K. H. Bengtsson\footnote{Retired from University of Borås, Sweden.\\e-mail: akhbengtsson@protonmail.com}}

\begin{document}

\maketitle

\begin{abstract}
This paper proposes a research program into how the special properties of the cubic chiral higher spin gravity theory may play a role in conformal cyclic cosmology and black hole physics, in particular for the black hole information paradox. The paper is qualitative, raising a number of questions and connections, that may be interesting to pursue in more detail. The proposal is based on a new -- at least not so commonly expressed -- view on the presumptive role of higher spin gauge fields in fundamental physics. In short, it is proposed that the non-unitary time evolution of the cubic chiral higher spin gravity theory is precisely what is needed for bridging the very large entropy gap between aeons in Roger Penrose's cosmological theory CCC.
\end{abstract}


\maketitle

\pagebreak

\section{Introduction}\label{sec:Introduction}
The content of this paper and the proposal it outlines is the confluence of quite a few ideas. Although none of them are perhaps not very new or original, to join them in a reasonable and comprehensible way, it seems necessary to deviate somewhat from the conventional synopsis of papers. Therefore, I will start by stating what is the driving force behind the deliberations to follow.

In my opinion there are two crucial, and related, questions regarding higher spin gauge fields, if one hopes or believes that they may play a role in fundamental physics: Where are they, that is, in what physical contexts do they appear, and once there, what role do they play?

The aim of the present paper is to offer an answer to these questions, or rather, a proposal for such an answer. I suggest that higher spin fields appear in extreme physical situations such as in black holes and that they play a role in resolving the black hole information paradox. Therefore they may also play a role as an ingredient in Penrose's Conformal Cyclic Cosmology.

The next section, number \ref{sec:ProposalOutline}, states the proposal in outline in order to get a first grip on what we are aiming for. The subsequent sections elaborate to some degree on the contributing ideas. Thus, Section \ref{sec:HistoricalBackground} is a brief history of the theory of higher spin gauge fields, emphasizing the question of the role to be played, if any, by these fields in fundamental physics. Section \ref{sec:CubicChiralTheory} describes the cubic chiral higher spin theory that is central to the proposal.  The conformal cyclic cosmology is described in Section \ref{sec:ConformalCyclicCosmologyEntropy}, in particular the question of entropy that is one problem that is driving the present proposal. Sections \ref{sec:HISGRA} and \ref{sec:CubicTheoryGeneralRelativity} dicuss some theoretical issues raised by the existence of the cubic higher spin gravity theory in relation to ordinary Einstein gravity. In the last section, number  \ref{sec:ChiralHigherSpinTheoryProvide}, I return to the proposal in more detail. 

\section{The proposal in outline}\label{sec:ProposalOutline}
Here we assume that the reader has a basic familiarity with the Penrose CCC (Conformal Cyclic Cosmology) theory. It will be described in Section \ref{sec:ConformalCyclicCosmologyEntropy}. At the core of this theory is the conundrum of the second law of thermodynamics in cosmology. Briefly stated, the entropy of matter and radiation has been increasing steadily since the Big Bang. Still it is assumed that the initial state of the universe was random, that is, a high entropy state. Indeed, the cosmic microwave background CMB of last scattering is thermal to a very high degree. However, this excludes the gravitational degrees of freedom. Gravity tends to clump matter, thereby increasing entropy immensely, easily svamping the matter and radiation contribution to entropy. Thus the total entropy of matter, radiation and gravitational degrees of freedom do indeed satisfy the second law. It is assumed that the gravitational degrees of freedom are not activated at the Big Bang (the so called Weyl curvature hypothesis of Penrose). Presumably, the universe ends in a state of many black holes of various sizes, some of them very massive, slowly evaporating into Hawking radiation. 

Here we reach a ``branch point'' in opinions. Stephen Hawking was initially of the opinion that in the process of evaporating, the entropy of a black hole was itself ``evaporating'', that is to say, it was decreasing in apparent conflict with the Second Law, information being lost \cite{HawkingYoung}. Later he changed his mind \cite{HawkingOld}. It seems that the majority of physicists agree with the ``later'' Hawking. Not so Penrose, however. In order for a smooth transition from an earlier high entropy aeon into a new low entropy aeon, information must be destroyed.

If one holds this point of view, it seems that one needs to come up with some physical process that actually can decrease entropy during black hole evaporation. As far as I know of, no such processes are known to exist among conventional accepted dynamical theories.\footnote{There is no preferred direction of time in the microscopical dynamical theories of verified physics.} Penrose has pointed out that the measurement process of quantum mechanics is such a process and that it is somehow related to quantum gravity, thereby rendering it dynamical.\footnote{See chapter 29 in the book \cite{PenroseRoadToReality}.} I cannot, and will not, add anything to that. Instead the suggestion offered here is that higher spin gauge fields may play the role.

There are nowadays a few theories of higher spin and versions thereof\footnote{See further discussion in Sections \ref{sec:HistoricalBackground} and \ref{sec:HISGRA} of the present paper.}, but we shall here focus on the one that is arguably the simplest and most ``conventional'' (if that characterization applies). This is the cubic chiral higher spin gravity theory in four space-time dimension. It is a complete theory involving fields of all (integer) spin with cubic interactions only.\footnote{It is based on the initial discovery of the cubic higher spin interactions in the light-front formulation in the early and mid 1980's by I. Bengtsson, L. Brink, N. Linden and the present author, and elaborated at the quartic level in the early 1990's by R. Metsaev. It was made fully explicit in 2016 by D. Ponomarev and E. Skvortsov. See more historical details and references in Section \ref{sec:HistoricalBackground}.} In the light-front formulation it is chiral, meaning that is is non-unitary (the Hamiltonian is not Hermitean), hence inducing processes that do not preserve probability. But this is precisely what we need. 

Thus the core of the proposal can be stated as a relation between the spin of gauge fields and their physical properties vis-\`a-vis entropy. For particles and fields with spin $\leq1$, those within the Standard Model of particle physics making up matter and electromagnetic radiation, time evolution tends to increase entropy corresponding to randomization. For spin $2$, time evolution tends to increase entropy corresponding to clumping of matter and radiation into ever more dense conglomerations, ultimately into black holes. Higher spin gauge fields, according to this proposal, are collectively responsible for decreasing entropy at very high energy, in a way that will be discussed below. The basic idea is the complete removal (from the phase space) of higher spin degrees of freedom excited within black holes. 

Thus stated, the proposal is bit vague, but it should convey the idea and the philosophy behind it. In the last section, it will be explained how the interactions of the cubic chiral higher spin theory may indeed transfer degrees of freedom from lower helicity to higher helicity, thus rendering them invisible to the world outside black holes. Higher spin gauge fields, that hitherto has been regarded as having no place in fundamental physics, do accordingly play a very specific role.

\section{Historical background on higher spin}\label{sec:HistoricalBackground}
Although the history of the main steps in the development of higher spin field theory is well known among the experts, it is perhaps appropriate to reiterate them here, as it has a bearing on the ideas underlying the proposal.

The theory of higher spin fields and particles is over 90 years old, if its birth is taken to coincide with the 1932 Majorana paper \cite{Majorana1932} followed by work of Dirac a few years later \cite{Dirac1936a}. For the first 45 years, the focus was mainly on massive fields. In particular so in the era of exploration of strong interaction physics where higher spin massive excitations occurred. The focus shifted to massless gauge fields with the work of Fronsdal in the mid 1970's \cite{Fronsdal1978}. The interest was enhanced in connection with research into supergravity in the late 1970's. Then in the early and mid 1980's, new attempts at finding theories for interacting higher spin gauge fields were met with some initial succes, first in the light-front gauge \cite{BBB1983a,BBL1987}, then covariantly \cite{BerendsBurgersvanDam1984,BerendsBurgersvanDam1985} and then also in AdS space-time \cite{FradkinVasiliev1987Grav,FradkinVasiliev1987Cubic}. A detailed history of, and introduction to this subject can be found in a two-volume text book \cite{HSFTvol1,HSFTvol2}.

However, as is well known by the experts on the subject, the initial high hopes for a imminent generalization of the successful gauge theories for spin 1 and 2 to higher spin, were soon crushed. This could perhaps have been expected, given the no-go theorems and the problems discovered at the interaction level already by Fierz and Pauli in 1939 \cite{FierzPauli1939a}. The very simple and systematic form of the cubic interactions in the light-cone gauge did not seem to generalize in any way to quartic and higher order, and this line of research was essentially put on hold.\footnote{The topic of cubic interactions in the light-front formulation was pursued in the 1990's by R. Metsaev. See for instance \cite{Metsaev1997cubic,Metsaev1993a}. As for the Metsaev work on the quartic level, I will return to it in Section \ref{sec:CubicChiralTheory}. As for the quartic order for pure spin $s$, I. Bengtsson, M. Cederwall and O. Lindgren looked for clues by expanding the action of Einstein gravity to quartic order in the light-cone gauge \cite{BengtssonCederwallLindgren1983}. No pattern to be generalized could be discerned. Subsequently, S. Ananth has performed similar studies, even up to quintic order \cite{Ananth2008}.} It was also practically clear that the covariant cubic spin 3 interaction of Berends, Burgers and Van Dam could not be generalized in any tractable way to higher spin without more powerful techniques. One such technique was provided by the BRST approach \cite{OuvryStern1986a,AKHB1986a,AKHB1988} and the techniques developed in references \cite{ManvelyanMkrtchyanRuhl2009a,ManvelyanMkrtchyanRuhl2010a,ManvelyanMkrtchyanRuhl2010b,ManvelyanMkrtchyanRuhl2011a}. Furthermore, a general analysis of the problem of extending the spin 3 theory to higher order showed problems with the gauge algebra \cite{BerendsBurgersvanDam1985,BB1986b,AKHB1985}. Although more powerful field theory techniques (BRST-BV antifield theory) \cite{BarnichHenneaux1993a} were subsequently applied to the problem of extending the theories to all spin and higher order interactions \cite{BekaertBoulanger2005a,BekaertBoulangerCnockaertLeclercq2006a,BekaertBoulangerLeclercq2010a,BekaertBoulangerCnockaert2006a}, these attempts only confirmed and reaffirmed problems encountered by the more naive initial attempts. 

The third approach, that of the Vasiliev AdS theory, for many years offered a strikingly different picture.\footnote{This very well known theory is reviewed in many places, for instance in the reviews \cite{DidenkoSkvortsovReview,BekaertCnockaertIazeollaVasiliev2005}. See also chapters 7 and 8 in \cite{HSFTvol2}. } This approach, with its quite different philosophy and and technical machinery, actually dominated the subject throughout the 1990's until well into the 2010's, when problems were unearthed in a concerted effort by newly arrived researchers to understand the theory better. It seems that a common denominator for the problems encountered in all three major approaches\footnote{The Dirac program (light-front), the Fronsdal program (covariant Minkowski) and the Vasiliev program (AdS) in the terminology of \cite{HSFTvol2}.} to higher spin gauge field interactions, is the impossibility to maintain locality at the quartic (and higher) order. A few inroad references to this research are \cite{MaldacenaZhiboedov2013,GiombiYin2010a,BoulangerKesselSkvortsovTaronnaVvsF,RoibanTseytlin2017a,Ponomarev2018a,BoulangerPonomarevSkvortsovTaronna2013a,SleightTaronna2018a}.

It should be mentioned that the search for higher spin interactions was conducted in the shadow of powerful no-go results, such as the Weinberg low-energy theorem \cite{WeinbergSNoGoII}\footnote{The Weinberg no-go theorem is a spin-off (no pun intended) of a series of systematic papers on S-matrix theory.}, the Coleman-Mandula theorem \cite{ColemanMandula1967} and the Aragone-Deser theorem \cite{AragoneDeser1971,AragoneDeser1979a,AragoneDeser1980a}, to mention perhaps the most well known and serious results. It would take us to far afield to go deeper into this, instead the reader may find relevant comments in reference \cite{BekaertBoulangerSundellYesGo}. Suffice it here to note that the Weinberg and Coleman-Mandula theorems concern the S-matrix, while the Aragone-Deser theorem is within field theory concerning the explicit coupling of higher spin gauge fields to gravity. The modern view, within the higher spin community, is not to try to circumvent the no-go theorems, but rather to accommodate them. We will see how this is achieved by the cubic chiral theory in an interesting way.

So what is the collective outcome of all these efforts during the last 50 years? To put it bluntly: there is no conventional (local and unitary) field theory for massless higher spin gauge fields in the physical dimension 4 of space-time.\footnote{Nor in any higher dimension.} Faced with this, the majority conclusion is to say that fundamental nature stops with spin 2. Theory tells us that we cannot, and should not, go beyond spin 2. Obviously this is not the point of view of the higher spin community of researchers. Instead there has always been the hope that  some higher spin theory will play a role in nature, presumably in some very high energy regime, perhaps in some unification scheme, while also contributing to solving the quantum gravity riddle. Even barring any such ``unrealistic'' hopes, it is certainly so that the difficulty of the higher spin interaction problem has intrigued the higher spin researchers and has been a strong motivating force.

In actual fact, much of the research into higher spin interactions was conducted not just in the shadow of the the no-go's, but often just ignoring them, in the hope that they could be evaded by some (hidden) assumption not being applicable to higher spin. As already noted, the modern view among the higher spin research community is instead to ``accommodate'' the no-go's. Thus the results of the last ten years of research into, in particular, light-front higher spin theory, is not viewed as having disproved the applicability of the no-go theorems, but rather made it clear how interacting higher spin fields may ``fly below the radar'' if the expression is allowed. The consensus of the workers in the field, is that the cubic chiral theory is consistent with the no-go theorems. I will explain how I understand how that is possible without rendering the theory without interest. There are certainly subtleties involved.

Therefore, in this article, I am suggesting such a role for higher spin fields that is somewhat unconventional. It is based on taking the cubic chiral higher spin theory seriously and asking the question of what it can be used for, of what role it can play in fundamental physics. Let us therefore continue with a description of the theory itself and some of its relevant properties.

\section{The cubic chiral higher spin theory}\label{sec:CubicChiralTheory}
The cubic chiral higher spin theory goes back to the very first positive results on higher spin self-interactions in 1983. In the paper \cite{BBB1983a}, cubic self-interaction terms were derived for arbitrary spin, i.e., interaction terms where three fields with arbitrary spin $s$ interacted in the pattern $s$-$s$-$s$. The result was derived in the light-front form of dynamics, and can be seen as an instance of the Dirac research program into all possible non-linear realizations of the Poincar{\'e} algebra \cite{Dirac1949FormsRelDyn} (the Dirac Program mentioned in footnote 7 above). It wasn't thought of in that way at the time, but this is a way of putting the paper in a wider context. 

A few years later, the theory was reformulated in a light-front string-like formalism using vertex operators \cite{BBL1987}. More important than the formalism itself, was the derivation of all possible cubic interaction terms between three fields of arbitrary spin in the pattern $s_1$-$s_2$-$s_3$.  A characteristic feature of the theory is that it is higher derivative, for instance for cubic spin $s$ self-interaction the number of derivatives in the vertex is $s$, and correspondingly for the mixed spin cases. Although it wasn't stressed at the time, this higher spin theory contains two-derivative interactions between spin 2 and higher spin, in other words, cubic gravitational interactions of higher spin.\footnote{As first pointed out in \cite{AKHB2014a}.} This does not contradict the Aragone-Deser no-go theorem, rather it goes to show that results on field theoretical coupling between fields are dependent on how the fields are represented.

Another few years later, R. Metsaev studied the quartic level of interactions for this particular theory \cite{MetsaevQuartic1,MetsaevQuartic2}.  An expression for a general quartic vertex was obtained, but it was formal in the sense that it involved the inverse of the free light-front Hamiltonian $h$. Therefore it was not well defined, but rather it indicated a non-locality at the quartic level.

In the light of this depressing situation, it is quite remarkable that the non-locality problem can be circumvented in the light-front theory. Buried in the very same Metsaev papers cited here, is the fact that there is a sub-theory that does not need any quartic interaction terms at all.

The light-front form of field theory only employs physical degrees of freedom. No auxiliary fields of any kind are needed or used. This means that any massless field of any spin $s$ only sports two field degrees of freedom in four-dimensional space-time, corresponding to the two helicities $\lambda=\pm s$. Consider a massless gauge field such as the photon. On the light-front, its four components are split up into $\varphi^+$, $\varphi^-$ and $\varphi^i$ with $i=1, 2$. The component $\varphi^+$ is gauged to zero, upon which the component $\varphi^-$ can be solved for in terms of the two transverse field components  $\varphi^i$. When this is properly done, no gauge freedom is left.\footnote{There is in actual fact a residual gauge invariance, pointed out by L. Brink, but it cannot be used for gauging away degrees of freedom.} What remains is two transverse degrees of freedom. The same kind of analysis can be performed for any massless higher spin field.

Now, for many related reasons\footnote{Not just computational, but also group theoretical, twistor theory related, and amplitude theory related reasons.}, it is natural to express the transverse field components of a momentum space spin $s$ field $\Phi(s, p_\mu)$ as a complex field $\varphi$ and its complex conjugate $\bar{\varphi}$ in the obvious way. In such a formalism $\varphi$ corresponds to helicity $\lambda=s$ and $\bar{\varphi}$ to helicity $\lambda=-s$. We may write $\varphi_s$ and $\bar{\varphi}_s$ for the positive and negative helicity components or as $\Phi_s=\varphi_s$ and $\Phi_{-s}={\bar\varphi}_s$ respectively.

This complex number rewriting of the theory is of course done also for coordinates, derivatives and in particular for transverse momenta. Indeed, $p$ and $\bar{p}$ carry orbital angular momenta $+1$ and $-1$ respectively. 

When the cubic interaction terms of the light-front theory is worked out, they all have the structure
\begin{equation}\label{eq:CubicVertex1}
C(\lambda_1,\lambda_2,\lambda_3)Y(p^+_1,p^+_2,p^+_3)\mathbb{P}^m\mathbb{\bar{P}}^n\Phi_{\lambda_1}\Phi_{\lambda_2}\Phi_{\lambda_3}+\text{c.c.}
\end{equation}
Here, $C$ stands for coupling constants which we will discuss below. $Y$ are rational functions of the $+$ direction momenta $p^+$ of the fields, conventionally denoted by $\beta$ so that $\beta=p^+$. The $Y$ functions were computed in \cite{BBL1987} (see formula \ref{eq:Yfunctions} below). 

The $\mathbb{P}$ and $\mathbb{\bar{P}}$ are certain combinations of transverse momenta $p_a$ and $\bar{p}_a$ and the $p^+_a$ momenta for the three fields entering the vertex (indexed by $a=1,2,3$), rendering the kinematic Lorentz invariance manifest. They carry orbital angular momenta $+1$ and $-1$ respectively (the $p^+$ carry no orbital angular momenta).\footnote{For details, se for instance chapter 6 och \cite{HSFTvol2}.} Since the Hamiltonian carries no total angular momentum, we have the formula
\begin{equation}\label{eq:TotalAngularMomentum1}
\lambda_1+\lambda_2+\lambda_3=n-m
\end{equation}
However, since cubic interactions involving powers of $\mathbb{P\bar{P}}$ are field redefinitions of the free Hamiltonian $h$ that contains a factor $\mathbb{P\bar{P}}$, we need only consider net non-negative powers of either $\mathbb{P}$ or $\mathbb{\bar{P}}$. One may therefore factor out $\mathbb{P}^m\mathbb{\bar{P}}^m$  and consequently only consider the terms
\begin{equation}\label{eq:CubicVertex2}
C(\lambda_1,\lambda_2,\lambda_3)Y(\beta_1,\beta_2,\beta_3)\mathbb{\bar{P}}^{\lambda_{123}}\Phi_{\lambda_1}\Phi_{\lambda_2}\Phi_{\lambda_3}+\text{c.c.}
\end{equation}
where
\begin{equation}\label{eq:TotalAngularMomentum2}
\lambda_{123}=\lambda_1+\lambda_2+\lambda_3
\end{equation}
These cubic interaction terms comprise the full list of possible interactions computed in \cite{BBL1987} with the functions
\begin{equation}\label{eq:Yfunctions}
Y(\beta_1,\beta_2,\beta_3)=\frac{1}{{\beta_1}^{\lambda_1}{\beta_2}^{\lambda_2}{\beta_3}^{\lambda_3}}
\end{equation}

Now, in order to have real Hamiltonian in the classical theory, or a Hermitean in the quantum theory, both the term written out in formula \eqref{eq:CubicVertex2} and its complex conjugate (indicated by c.c.) must be included. However, if one anyway just takes either of the two conjugated terms, one may have a theory that do not require any quartic completion, provided that the otherwise arbitrary coupling constants $C(s_1,s_2,s_3)$ have a particular form, first determined by Metsaev in \cite{MetsaevQuartic1,MetsaevQuartic2}. This fact was however not appreciated at the time in the early 1990's.

It wasn't until the whole setup was recomputed in 2016 by D. Ponomarev and E. Skvortsov \cite{PonomarevSkvortsov2016a} that it became clear that there is a purely cubic higher spin field theory provided that the coupling coefficients are chosen according to 
\begin{equation}\label{eq:MetsaevCoefficients}
C(\lambda_1,\lambda_2,\lambda_3)=\frac{l_p^{(\lambda_1+\lambda_2+\lambda_3-1)}}{\Gamma(\lambda_1+\lambda_2+\lambda_3)}
\end{equation}
and that the c.c. terms in the Hamiltonian are dropped. The price to pay is the non-unitarity of the theory. In formula \eqref{eq:MetsaevCoefficients}, $l_p$ is a coupling constant of length dimension such as the Planck constant. More on that in the last section.

In the subsequent years since its explicit formulation, the theory has been developed in several ways, two of which is covariantization and exploration of its quantum properties.

\paragraph*{Covariantization}

It should not be expected that the theory can be covariantized using ordinary Lorentz tensors for the covariant higher spin gauge fields, as the no-go results are numerous and well-established. Instead the procedure is based on twistor theory \cite{KrasnovSkvortsovTranCovI,SkvortsovVanDongenCovIIa,SharapovSkvortsovSukhanovVanDongenCovIIb}\footnote{This ongoing work but I confine myself to citing the initial results as this is not in the main direction of the present paper. }, or rather, on two-component spinor language as introduced in the book \cite{PenroseRindlerVol1} and developed by several authors (for references, see the papers cited here). It would take us to far afield to enter into this fascinating subject.

\paragraph*{Quantum properties}

As for quantum properties, there is a series of initial papers investigating the theory. The first one is a letter \cite{SkvortsovTranTsulaiaQI}, reporting the following three results: 

\emph{First}, due to the particular form of the three-point couplings, all on-shell tree amplitudes vanish, and the $S$-matrix is $1$. As the authors point out, this makes the theory consistent with the no-go theorems at least at the tree level. 

\emph{Second}, all vacuum diagrams vanish. For the one-loop ``bubble'' diagram, this hinges on the need to regularize the total number of degrees of freedom. This sum is infinite, but may be regularized to zero using zeta function regularization \cite{GiombiKlebanov2013,BeccariaTseytlin2015a}. According to the paper, all other vacuum diagrams vanish without the need to regularize, instead relying on the particular form of the coupling factors. Such zeta-function regularization is not uncommon in theories with infinite number of states. 

\emph{Third}, the paper argues that all loop diagrams with external legs vanish given that the total number of degrees of freedom is regularized to zero.

The second paper \cite{SkvortsovTranTsulaiaQII} and third paper \cite{SkvortsovTranQIII}, provide much more detail and strengthens and extends the result the first paper. It thus seems that the cubic chiral theory is an UV-finite higher spin theory with a trivial $S$-matrix consistent with the Minkowski no-go theorems. All the same, there are actually interactions, gravitational as well, albeit cubic and non-unitary.

There is however something strange with these results. How can the S-matrix be a unit matrix (which is certainly a unitary matrix) and the underlying theory be non-unitary? As I will be arguing, the resolution of  this apparent discrepancy may actually provide the subtle mechanism needed in order for the non-unitary evaporation of black holes not to contradict the second law of thermodynamics. Technically, what is proved in the cited papers, is that the amplitudes (tree as well as loop) computed from the cubic Hamiltonian of the theory, vanish. These amplitudes, however, are not the full amplitudes that could have been calculated had the c.c.-terms in the cubic Hamiltonian \eqref{eq:CubicVertex1} been included. Then the theory would have been unitary but severely quartic non-local and probably non-polynomial. Something subtle is going on here that is most likely not fully understood at the moment. Some further comments on this will be added in Sections \ref{sec:HISGRA}, \ref{sec:CubicTheoryGeneralRelativity} and  \ref{sec:ChiralHigherSpinTheoryProvide}.

\section{Conformal cyclic cosmology and entropy}\label{sec:ConformalCyclicCosmologyEntropy}
Conformal Cyclic Cosmology is a cosmological theory put forward by Roger Penrose around 2005 \cite{PenroseOutrageous} and explicated in the book \cite{PenroseCCCBook} and in the article \cite{Penrose2014CCC} as well as in many talks and interviews. I will outline the main properties of the theory, without technical details, in order to present a context for the present proposal. For work on the mathematics of the theory and for evaluations of the predictions of the theory against observations the reader can consult references listed in the papers \cite{Eckstein2022,TodCCCpreprint,Tod2021minicourse}. Two recent papers are \cite{BodniaEtAl2024,MeissnerPenrose2025}. Whether the theory is consistent with observations, or ruled out, is still a contentious issue. I have nothing to add there. Our aim here is anyway not to discuss the pros and cons, theoretically and observationally, of the CCC theory. 

Suffice it to say that theory is clearly an alternative to the standard cosmological theory $\Lambda$CDM and that there is certainly other alternatives such as the theory of N. Turok and collaborators. See for instance \cite{BoyleFinnTurok2022a} which refers back to earlier publications within this program. Furthermore, new cosmological data are continuously incoming, not the least from the James Webb Telescope, so the existence of alternative theoretical schemes is certainly good from a general philosophy of science perspective.

Thus rather than arguing the merits and deficits of the CCC theory, instead assuming that the theory is a viable cosmological theory, the aim is to propose a positive, indeed essential, role for higher spin fields or degrees of freedom in cosmology. That is the aim of the present discussion.

It could be maintained that although Penrose himself designated the theory as ``outrageous'' it is actually a quite conventional alternative cosmological theory. It is not a multiverse theory and it has observational consequences that can be, and are, tested. According to CCC, there is just one universe, that however, goes through successive cycles -- called aeons -- joined by crossover surfaces, connecting the far future of one aeon to the big bang of the next. The metric of the universe, with its cosmic time, extends through the crossover surface. In order for this to be possible, there is an infinite conformal rescaling at the crossover (or rather infinite rescalings at each side of the crossover so that the very large late aeon universe can be joined to the very small new universe). Clearly this is a somewhat ``outrageous'' notion, and it is here that one finds problematic features of the theory from the theoretical side. Conformal rescaling requires that fields and particles are massless and thus stray massive and charged particles such as protons and electrons are a challenge to the scheme, unless they are all swallowed by black holes or decay in some way. Again, this problem is not the focus of the present proposal.

Central to the Penrose theory is the puzzle of the second law of thermodynamics. The beginning of an aeon is a big bang, a state of the newborn universe that is very low in entropy. Indeed, it is this low entropy that, through the second law of thermodynamics, can be seen to ``drive'', so to speak, the evolution of the universe. Had the universe started in a maximum entropy state, then not very much interesting would have been possible to happen subsequently.\footnote{As already alluded to, it is gravity that is responsible for ``this interesting to happen'' through the clumping of matter into stars, planets and galaxies and so on and so forth.} On the other hand, if the very high entropy ending of one aeon is to be connected (the geometry of which we will not describe here) to the low entropy big bang of the next, not just the geometrical problems must be accounted for by the cosmological theory, but also the role of the second law of thermodynamics. What seems to be needed is some resetting of the entropy at the border crossing between aeons or rather prior to it.

The second law is of course a fundamental law of nature, perhaps the most well established of them all, one that we can verify almost without doing anything in particular, just letting things run down without intervention. Processes in nature seem to be generally irreversible, unless some constructive agent intervenes to produce local order and positive development. A notable example being our own biosphere driven by the our local hot sun. But the price is increased disorder elsewhere in a larger context. All this is of course well known.

At the same time, theoretically, there is an awkward problem that every student of statistical mechanics is confronted with. All the microscopic laws of nature are time reversible which seems to be at odds with the time direction implied by the second law. In classical statistical mechanics the problem is ``solved'' by coarse graining.\footnote{I personally think one has to be quite tone-deaf as a student if one does not read between the lines quite some unease on the authors parts, or the lecturers part for that matter.} Even though the equations of motion may be run in the reverse order, the natural direction of time evolution is towards larger and larger coarse grained regions in phase space, corresponding to more random looking macroscopic situations.  In quantum statistical theory, the situation is a little less ad hoc in that coarse graining the continuous classical phase space can, at least in principle, be replaced by counting all the discrete quantum states corresponding to a given macroscopic situation.\footnote{Very eloquently explained in the textbook \cite{KittelThermalPhysics}.}

In quantum mechanics, time reversibility is represented by unitary evolution. This is an almost sacred aspect of quantum mechanics, yet in practice,  the theory does not work without the non-unitary process of measurement. Now this is an area of much agony in theoretical physics. For our purposes here and for now, it is enough to state that the dichotomy is there and does not seem to have any simple solution that will satisfy all inclinations. Perhaps some amount of irreversibility in the microscopic laws of nature may not be such a bad thing after all.

Coming back to CCC, Penrose argues that the late stages of an aeon is dominated by big black holes, such as have been residing in the centers of galaxies of merging galactic clusters. As the background temperature drops below the black hole surface temperatures they will evaporate through (thermal) radiation. Whether the radiation is actually thermal or not, constitutes the black hole information paradox. Most of the entropy of the late universe will reside in such black holes.

Thus it seems that CCC requires a non-conventional stance regarding the so-called black hole information paradox.\footnote{See however \cite{Eckstein2022} for an alternative point of view.} This is a large field of research and again this is not the place to discuss the  various approaches to the problem, i.e., the problem of showing that information is indeed preserved in some way during black hole evaporation.\footnote{I despair of providing references here and contend myself the two Hawking papers cited in Section \ref{sec:ProposalOutline}. It is not difficult to find articles treating the question.} Instead, I will side with Penrose, who is siding with ``the early Hawking'', that maintained that information is actually lost in black hole evaporation. We will now see how the cubic chiral higher spin gravity theory may play a role in such a scenario.

\section{The conundrum of higher spin gravity}\label{sec:HISGRA}

It has become customary to refer to higher spin gauge theory as "higher spin gravity" abbreviated by HiSGRA, the reason being that everything must interact with gravity, and so must the higher spin gauge fields do if they exist. Such HiSGRA theories are rare. In three space-time dimension there are Chern-Simons-type theories dating back to 1989 \cite{Blencowe1989,PopeTownsend1989,BergshoeffBlencoweStelle1990,FradkinLinetsky1990}. There is also a conformal HiSGRA in four dimensions extending conformal Weyl gravity \cite{Segal2003ConformalHS}.\footnote{For more references and a brief discussion, see reference \cite{KrasnovSkvortsovTranCovI}.} However, the cubic chiral theory seems to be the only theory with interactions, propagating degrees of freedom and an action\footnote{The free part of which, turning off gravity and self-interactions, just describes massless particles, or propagating waves, of (integer) helicity $\lambda=\pm s$.}, albeit non-unitary. To its advantage can also be added that it can be constructed, and was indeed discovered, using a minimum of technical machinery.

Coming back to the question of gravitational interactions, this leads to a severe problem. The spin 2 part of gravity interacting with everything else is a highly non-linear theory, the pure spin 2 part described by Einstein gravity. The cubic chiral higher spin theory -- including a spin 2 field -- only involves cubic interactions. Faced with this, most researchers in the area seem to hope that the cubic theory will eventually be a part of a fully interacting, and unitary, theory including spin 2 gravity. To be relevant to our world, there would also have to come about some unification with the actual lower spin interactions of fundamental matter and radiation physics. 

The present proposal certainly seems to be in conflict with such a scenario as it presupposes that higher spin effectively stops at the cubic chiral theory. Of course, a unification with the particles and fields of fundamental matter and radiation physics is necessary also in this scenario. But it may entail a drastic rethinking of quantum gravity. To get a grip on this problem, let us return to the essence of the no-go theorems.

\section{The cubic theory and general relativity}\label{sec:CubicTheoryGeneralRelativity}

It is a very well established fact that starting from a free spin 2 field and attempting to self-couple it, one will end up with a non-polynomial theory that is actually Einstein gravity.\footnote{The general systematic procedure is now called ``Noether coupling''.} This program, sometimes called the ``Gupta Program'' after the first paper on the subject \cite{Gupta1954}, took a long time to establish in practice with contributions from many authors: R.H. Kraichnan in 1955 \cite{Kraichnan1955}, W.E. Thirring in 1961 \cite{Thirring1961}, R.P. Feynman in 1962 \cite{Feynman1962a}, W. Wyss in 1965 \cite{Wyss1965}, S. Deser in 1970 \cite{Deser1970} and D.G Boulware and S. Deser in 1975 \cite{BoulwareDeser1975}. The history of the program is described by Fang and Fronsdal in their paper \cite{FangFronsdal1979} where the name of the program is also coined. Fang and Fronsdal also proposed a ``generalized Gupta program''\footnote{Now called the ``Fronsdal program''.} to do the same for higher spin gauge fields. This however, as we now know, was a problem on a completely different scale of difficulty. Fronsdal soon realized that in order to have any hope of success one would need to include fields of all (integer) spin \cite{Fronsdal1979conf}.

The problem was not only very difficult, it turned out to be impossible to solve the higher spin gauge interaction problem, the history of which we have briefly outlined in Section \ref{sec:HistoricalBackground} above, except in some very few special cases, one of which is the cubic chiral theory.

This raises a question. There are interactions in the cubic chiral theory between spin 2 and higher spin and between spin 2 with itself. These cubic interactions, in particular the cubic spin 2 self-interactions, shouldn't they seed the full tower of higher order spin 2 self-interactions, thus leading again to full Einstein gravity coupled to a higher spin gauge theory involving all spin and presumably also non-polynomial? 

However, such a theory does not exist. It is impossible to construct within covariant Minkowski gauge field theory, as has been shown in the work referred to above in Section \ref{sec:HistoricalBackground}. Then turning to the light-front physical field formulation, the program leads to severe non-localities at the quartic level, confirming the results of the covariant Fronsdal approach. Except, there is an escape exit, namely the cubic chiral theory, apparently impossible to see in the covariant Fronsdal program.\footnote{Although history did not take that route, the theory could, perhaps, in principle have been discovered using twistor inspired methods, had they been applied to the problem.} 

The situation seems to be the following: For pure massless spin 2, the Noether coupling procedure leads to the non-polynomial theory of Einstein gravity. For a theory with massless fields of all integer spin including spin 2, the procedure leads to the cubic chiral theory. Thus we have two theories of gravity, one standard Einstein gravity, the other the cubic chiral HiSGRA.

Faced with this curious state of affairs, most researchers seem to hold the view that the cubic chiral theory will eventually be part of a fully interacting unitary gravitational theory. In my opinion that is highly unlikely since it is precisely what has already been ruled out. If such a theory exists, shouldn't the Dirac light-front physical field procedure of closing the Poincaré algebra see it?

But as evidenced by the appearance of new papers on the theory, see for instance the recent \cite{serrani2025classification,tran2025chiral,tran2025anomaly,masonsharma2025chiral,ponomarev2025chiral}, there has already appeared, and will undoubtedly appear new, interesting versions of the basic cubic chiral HiSGRA. Of course, higher spin theory has surprised us before, and there may be another escape exit, remaining to be discovered, where the higher spin part of a presumptive theory is cubic while the spin 2 part sums to Einstein gravity. Instead of hoping for such a theory to appear, I will assume that higher spin effectively stops with the cubic chiral theory adapted to black hole physics.

\section{What a non-unitary cubic higher spin theory may provide}\label{sec:ChiralHigherSpinTheoryProvide}
Let me sum up the arguments. The ideas put forward here might seem radical, but they can also be viewed as quite conservative. There are radical points, one of which is the non-conventional hypothesis that higher spin gauge fields do play a role in fundamental physics and taking the cubic chiral theory seriously as a quantum HiSGRA theory. We will see that following through on the logic of the proposal leads to difficult questions that, however, should be possible to address within ordinary field theory.

The proposal is conservative in that it accepts the lessons from 50 years of research into higher spin gauge interactions. That is, the only interacting higher spin gravity theory is cubic and non-unitary. This theory is actually quite conventional in that it is easy to write down, it has -- as already pointed out -- propagating degrees of freedom, an action, and it resides in physical 4-dimensional space-time. The theory may be there for a reason. 

The pure gauge field part of it is well known and very simple in the light-front formulation (see Section \ref{sec:CubicChiralTheory}). It remains to merge it with the Standard Model particles and fields (or perhaps a modified version thereof) if it is to play a fundamental role. Needless to say, such a theory would be relevant only at very high energy and in extreme situations such as in the interior of black holes. This, however, leads to a serious conundrum, as noted in the previous section,  in that the cubic nature of the interactions -- gravitational included (or rather spin 2) -- clashes with the non-polynomial nature of the Einstein gravitational interactions. To face this problematic issue, there is a radical point, namely the hypothesis that as higher and higher energies are investigated, in certain physical situations such as in black holes, the Standard Model coupled to Einstein general relativity transforms into a "standard model" coupled to a quantized cubic higher spin theory. This may seem like a wild speculation, and perhaps it is, and there are certainly many objections that can be raised. But there is also a conservative aspect to it. Research into quantum gravity is almost as old as Einstein gravity itself and it has only met with limited success  despite the efforts of many workers. Indeed, a hundred years after the advent of quantum theory, one might well claim that gravity has resisted all attempts at quantization.

At the root of the quantum gravity problems is the highly non-linear nature of classical gravity. Quantum mechanics works best with linear systems where the interactions can be treated as perturbations as in the Standard Model of particle physics. Indeed, the basic interactions of the Standard Model are cubic, supplemented with quartic interactions for gauge invariance consistency. Now the pure cubic chiral theory is of course only cubic, which in respect to quantum mechanics is a good feature.

So let us then come back to the entropy issue. I will loosely talk about ``outside'' and ``inside'' black holes without specifying what is exactly meant by that, as it is premature at this stage, although it may be the horizon.\footnote{If there is some kind of transformation from standard general relativity into a cubic higher spin theory, that may affect the precise location and nature of the horizon, but most likely an ``inside'' and an ``outside'' of the black holes would still be a meaningful notion. } Thus outside black holes we have Einstein gravity coupled to matter and radiation in the standard way.\footnote{I'm clearly dodging the issue of whether one would not in principle need quantum gravity everywhere, not just inside black holes. In practice it seems that we can understand quite dense astronomical objects such as stars, white dwarfs and neutron stars without a full quantum gravity theory.} Here there cannot be any higher spin gauge fields. They cannot even propagate as free waves since everything must interact with gravity, and that is not possible according to the coupling no-go theorems of the Aragone-Deser type. So outside black holes there aren't any interacting higher spin fields, nor any free. That is, no higher spin degrees of freedom of any sort.\footnote{This may seem like a funny way of phrasing it, but I hope the reader gets the point. For all practical purposes, free fields doesn't exist except as mathematical entities. Things that do not interact cannot be detected.} This is also a consequence of the Weinberg low energy theorem and the Coleman-Mandula no-go theorem. There are no detectable long range effects of higher spin fields. The cubic chiral theory is consistent with this, having a trivial unit S-matrix. 

Inside black holes, one may consider at least two scenarios. Either Einstein gravity and what may remain of the Standard Model are coupled to a non-unitary higher spin theory. Or Einstein gravity is replaced by a cubic chiral higher spin theory, including spin 2 needless to say, coupled to matter and spin 1 force fields. In both scenarios, the lower spin degrees of freedom transitions into higher spin degrees of freedom via the cubic interactions. The S-matrix however being trivial, these do not subsequently scatter, although there may be transitions according to the cubic interactions. I will return to this question at the end and be a little more explicit regarding the transitions from lower spin to higher spin.

The paradox of higher spin existence is solved by a reduction of phase space, or the space of quantum states, in the sense that the part of phase space containing higher spin states factors out. The degrees of freedom in there cannot interact with the rest of the states, except possibly in a very feeble way to be discussed below. For all practical purposes, they do not exist. Entropy in this way resets to a low value. In other words, once a lower spin degree of freedom has transitioned into a higher spin degree of freedom, it is effectively lost to the phase space. This is is agreement with the way Penrose pictures the process in \cite{Penrose2014CCC}, a ``renormalization'' of entropy as it were. Indeed, one should note that Penrose stresses that the resetting of entropy should not affect local physics, as that would violate the second law of thermodynamics. Instead, it should be global process in the form of a reduction of phase space. This is what a proper application of the cubic chiral theory may provide. Also as pointed out by Penrose, the resetting of entropy need not be a sudden process, it may occur gradually throughout the black hole evaporation process.

I'm aware of a serious problem with this scenario that I haven't yet touched upon. The cubic chiral theory, isn't that derived within Minkowski, or possibly a constant curvature, space-time? How can it be relevant inside black holes? The question cannot be evaded by suggesting that the higher spin theory should be derived on a general relativistic background, since we already know that that is not possible. Here, I think, one must follow through on the logic of the argument to its conclusion. Yes, inside black holes, space-time may be simple. In order for the scenario painted here to be realistic, it is clear that some serious rethinking is needed. However, it may not be such a bad thing after all. A transformation from ordinary gravity outside black holes to a  cubic HiSGRA inside black holes, while preserving a horizon, may be another virtue of the scheme rather than a vice. One would have a simpler theory to work with inside black holes.

However, there is still the question of the transformation from physics outside a black hole to physics inside it. Is it a sudden shift, where although the curvature itself can be continuous, it is not differentiable? Or is the shift smoothed by quantum effects? Clearly I cannot answer these questions, but I can offer some more thinking. 

I have written loosely about the cubic chiral HiSGRA unified with some modified standard model of matter particles and spin 1 gauge fields being the theory reigning inside black holes. As for the spin 1 gauge fields they will carry the gauge group of low energy physics, perhaps unified in some version of a GUT gauge group with coupling parameter $\alpha_g$. Indeed, looking back at the Metsaev coupling coefficients \eqref{eq:MetsaevCoefficients}, the coupling parameter should really read
\begin{equation}\label{eq:HSCouplingParameters}
 \alpha_gl_p^{(\lambda_1+\lambda_2+\lambda_3-1)}
 \end{equation}
in order for a pure $1$-$1$-$1$ interaction to have coupling $\alpha_g$ and a pure $2$-$2$-$2$ interaction to have coupling $\alpha_gl_p=L_P$ where $L_P$ is the Planck constant. Then a pure spin $s$-$s$-$s$ self-interaction comes with the coupling $\alpha_g(L_P/\alpha_g)^{(s-1)}$. Thus higher spin self-interactions are suppressed by factors of $L_P/\alpha_g$ as compared to the lower spin interactions Perhaps this simple observation might offer some smoothing of the transition between theories inside and outside black holes, possible augmented by considerations of how the couplings may run with energy.

As a final item, let me return to and try to be somewhat more concrete, at least qualitatively, on the transitions from lower spin to higher spin, which is crucial for the reduction of the phase space discussed above. We want lower spin degrees of freedom to transition into higher spin degrees of freedom. The interactions being cubic, it is natural view them in analogy with quantum electrodynamic pair production. 

First note that for a general $s_1$-$s_2$-$s_3$ vertex the coupling parameter is $\alpha_g(L_P/\alpha_g)^{\lambda_{123}-1}$, the helicities being $\lambda_1=s_1,\lambda_2=s_2,\lambda_3=-s_3$ or some permutation thereof, with $\lambda_{123}=\lambda_1+\lambda_2+\lambda_3$.

Consider for instance a vertex with $\lambda_1=1,\lambda_2=s,\lambda_3=-s$ corresponding to a helicity 1 field creating a helicity $\pm s$ pair. Then $\lambda_{123}=1$ and the coupling parameter is simply $\alpha_g$ and the transverse momentum factor is $\mathbb{\bar{P}}$ regardless of the higher spin helicities. 

Of course, there will transitions back from higher helicity into lower helicity. Consider for instance a helicity $s$ field creating a helicity $\pm 1$ pair. Then we have $\lambda_1=s,\lambda_2=1,\lambda_3=-1$ and $\lambda_{123}=s-1$. The coupling parameter becomes $\alpha_g(L_P/\alpha_g)^{(s-2)}$. For helicity $s=3$ we thus get coupling parameter $L_P$ and the transverse momentum factor is $\mathbb{\bar{P}}^2$. The transitions back from higher helicity to lower helicity are therefore strongly suppressed by factors of $L_P/\alpha_g$. 

Performing a similar simple calculation for spin 2 transitioning into, say spin 4 and higher even spin, shows that it is slower -- not unexpected -- by a factor of $L_P$, but transitions back from higher even spin to spin 2 is even slower by multiple factors of $L_P$. The cubic interactions of the theory interpreted in this way lends support to the hypothesis that the cubic chiral theory forces a gradual migration of lower spin degrees of freedom into higher spin degrees of freedom.

What more can be said in favor of the cubic chiral higher spin theory? There is of course no end to creativity when it comes to proposing mechanisms for explaining puzzles within theoretical high energy physics and cosmology. Nothing wrong with that. But the cubic chiral theory has perhaps one advantage as compared to more ad hoc suggestions. It is simple and it is based on first principles and the question: Can one construct an interacting field theory for higher spin gauge fields? The answer is yes, and if it is not to be a mere curiosity, the next question is what its role is. According to the present proposal, its role is a very specific one, namely to reset the entropy during black hole evaporation. Lower spin degrees of freedom transitioning into higher spin phase space never to be seen again.

\section{Concluding remarks}\label{sec:ConcludingRemarks}

I haven't tried to provide technical calculations supporting the proposal, being well aware of the fact that there are many more higher spin researchers much more qualified for this kind of work. Instead I hope to seed some interest for work in this direction. As for final words, if one hopes for some role to be played by higher spin in the larger scheme of things, then that role is very likely unconventional, as this area of research into theoretical physics has itself always been highly unconventional, driven as it has been by the severe constraints of consistency. 

\section{Acknowledgment}\label{sec:Acknowledgment}

I thank Bo Sundborg for numerous discussions on these questions over the last couple of years. The ideas presented here has been on my mind for quite a few years. Upon reading two recent higher spin papers \cite{serrani2025classification,tran2025anomaly} over the midsummer vacation, I was inspired to rethink it all and  writing it up.
\pagebreak

\end{document}